
\input phyzzx
\overfullrule 0pt
\hoffset=1cm
\hsize=5.7in  \vsize=8.75in
\parskip=0pt
\def\ie{{\it i.e.}}
\def\eg{{\it e.g.}}
\def\etal{{\it et al.}}

\def\unlock{\catcode`@=11} 
\def\ev{{\rm eV}}
\def\gev{{\rm GeV}}
\def\mev{{\rm MeV}}
\let\sun=\odot
\def\lock{\catcode`@=12} 
\itemsize=19pt
\unlock
\refindent=19pt
\def\refitem#1{\r@fitem{#1.}}
\def\refout{\par\penalty-400\vskip\chapterskip
   \spacecheck\referenceminspace
   \ifreferenceopen \Closeout\referencewrite \referenceopenfalse \fi
   \line{\hfil \bf REFERENCES\hfil}\vskip\headskip
   \input \jobname.refs
   }
\def\chapter#1{\par \penalty-300 \vskip\chapterskip
   \spacecheck\chapterminspace
   \chapterreset \centerline{\caps #1}
   \nobreak\vskip\headskip \penalty 30000
   {\pr@tect\wlog{\string\chapter\space \chapterlabel}} }
\lock
\sequentialequations
\chapterminspace=6pc
\sectionminspace=5pc
%
\Pubnum{SCIPP 93/31}
\date{October, 1993}
\pubtype{  }     
\titlepage
\vskip5cm
\singlespace
\centerline{{\fourteenbf
Particle Astrophysics from the Particle Perspective}
\foot{Work supported in part by the U.S. Department of Energy.}}
\vskip12pt
\centerline{\caps Michael Dine}
\vskip2pt
\centerline{\sl Santa Cruz Institute for Particle Physics}
\centerline{\sl University of California, Santa Cruz, CA 95064}
\vskip2cm
\vbox{%
\narrower  \tenpoint \baselineskip=12pt
}
\vfill
\centerline{Invited Talk}
\centerline{16th International Symposium on Lepton and
 Photon Interactions}
\centerline{Cornell Unviversity, Ithaca, NY, 10--14 August, 1993}
\vfill\endpage

\chapter{INTRODUCTION}

\REF\cobe{G.F. Smoot \etal, Ap.~J. {\bf 396} (1992) L1.}
During the last decade, particle astrophysics has developed
into a discipline by itself, encompassing a large
array of interesting phenomena.  I am certainly not an expert on
what we might refer to as the ``hard core" astrophysics aspects of
the subject, particularly those associated
with structure formation and the detailed interpretation of the COBE
data, or of detailed theories of inflation.  What I would like to do
today, after briefly surveying the standard big bang theory,
inflation, and the marvelous results from COBE,\refmark{\cobe}
is largely
to quote uncritically from the experts on such topics, and turn
to those aspects of the subject which have the most immediate implications for
present day particle physics.

\REF\hcdm{M. Davis, F.J. Summers and D. Schlegel, Nature {\bf 359}
(1992) 393; A.N. Taylor and M. Rowan-Robinson,
Nature {\bf 359} (1992) 396;
R.K. Schaefer and Q. Shafi, BA-92-28 (1992); J.A. Holtzman and
J.R. Primack
Ap.~J. {\bf 405} (1993) 428; A. Klypin, J. Holtzman, J.R. Primack
and E. Regos, Ap.~J. to appear.}
Even here, one could make a large list.  In my briefcase, I am
carrying literally
hundreds of papers, and if one followed their references,
the number would
surely extend into the thousands.  To narrow this list further, I would
like to focus in this talk on a few principle topics:
\item{1.}
Neutralino cold dark matter.
There are two particularly well-motivated
candidates from particle physics at the
present time.  The first of these, which has direct relevance for
accelerator experiments, is the neutralino, likely to be the
lightest new particle if supersymmetry is correct.  I will
review why this is such a wonderful dark matter candidate,
and some of the proposals for searching for this component
of the dark matter directly.  During the past year or so, a major
industry has developed trying to constrain the parameters of
supersymmetry -- the squark and gaugino masses, and so forth,
and the idea of supersymmetric dark matter has been a major
component of these efforts.  I will discuss briefly some of the
assumptions that go into these analyses and their plausibility;
this is also important to understanding the constraints set by
present and future direct dark matter searches.
\item{2.}
The second of these candidates is the axion.
The axion is  the subject of experimental searches of growing
sophistication and power. The situation, however, is qualitatively
different than that of the neutralino; if the axion is {\it not}
the dark matter, it is probably impossible to find.  Indeed, if
an axion is found, this will be particularly exciting because it
will provide a window on an energy scale that is otherwise
completely inaccessible.
I will review
the axion idea and the related experiments.  Recently, the
plausibility of the axion idea has been ``attacked;" I will review these
arguments, and give some counterarguments.
\item{3.}
Neutrinos.  The question of neutrino
mass is a long-standing one, and there are a variety
of theoretical ideas which suggest that one neutrino might
have a mass in the few eV range.  Experimentally, there
are a number of hints, of varying degrees of credibility,
of neutrino mass:  the solar neutrino
problem and the lower than expected flux of muon neutrinos
from cosmic rays being the most impressive.
Moreover, among astrophysicists and cosmologists, a model with a
mix of cold (70\%) and hot (30\%) dark matter has become very
popular in light of the COBE results.\refmark{\hcdm}
A light (\eg, 7 eV) neutrino
is an often-mentioned candidate for this dark matter.  These
subjects have been reviewed in the talk by Smirnov at this
meeting, and I will only say a few words about them here.
\item{4.}
Electroweak baryogenesis:  In the past few years, it has become
clear that
the observed baryon asymmetry may have been produced at the
electroweak phase transition.  If so, this might resolve a number of
questions
in particle astrophysics.  It also has bearing on extensions of the standard
model,
in particular on questions of $CP$-violation.
I will give a brief overview of
this subject here, and mention some efforts which
it has (in part) motivated to understand the problem of
$CP$-violation at high energy colliders.
\item{5.}
Exotica:  Under this heading, I will discuss a variety
of more speculative topics.  These include:  the fate of domain
walls; axion cosmology in the framework of supersymmetry
(axinos, saxions, etc.); cosmological constraints on models
of dynamical supersymmetry breaking; the possibility that much
of the dark matter is in the form of a very small cosmological
constant.

Perhaps before beginning, however, since this is at best a ``mixed
audience,"
it is worth recalling some of the highlights of the standard big bang
theory.  The big bang cosmology starts with the observation that, on the
average, the galaxies are all moving away from us at a rate
proportional to their distance, and with the ``cosmological principle,"
the idea that our position in the universe is in no sense special,
and that on very large scales, the universe is homogeneous and
isotropic.  There is strong observational evidence for this; the most dramatic
being
the COBE results on the fluctuations in the
microwave background temperature.
This, plus Einstein's equations, lead to
a model of space-time with metric of the form
$$ds^2 = -dt^2 + a^2(t)\left[{dr^2 \over 1 - kr^2} + r^2(d\theta^2
+\sin^2(\theta) d \phi^2)\right].\eqn\friedman$$
Here $a$ is the scale factor, and satisfies the equation
$$\left({\dot a \over a}\right)^2 ={8 \over 3} \pi G \rho - {k \over a^2}
\eqn\einstein$$
where $G$ is Newton's constant and $\rho$ is the energy density
of matter.  $k=\pm 1,0$.  The Hubble ``constant" is $H=\dot a/a$;
today, $H_o =100 h$ km s$^{-1}$ Mpc$^{-1}$.  The ``critical density"
is determined by the requirement that $k=0$ in eq.~\einstein:
$$\rho_{c}={3 H_o^2 \over 8 \pi G}= 1.054 \times
10^{-5} h^2~\gev~{\rm cm}^{-3}.\eqn\rhocrit$$
Extremely important in all of these quantities is the parameter
$\Omega={ \rho_o / \rho_c}$.

\REF\nucleosynthesis{See, \eg, T .P. Walker, G. Steigman,
D.N. Schram, K.A. Olive and H.S. Kang, Astrophys. J. {\bf 376} (1991)
51.}
\REF\knox{L. Knox and M. Turner, Phys. Rev. Lett. 70 (1993) 371.}
Now if one runs this picture backwards in time, matter becomes
more and more compressed, and the temperature rises.
In fact, we can think of the temperature as a sort of clock,
labeling the important moments in the history of the universe.
There are two particularly relevant periods:  that of radiation
domination and that of matter domination of the energy density.
Radiation domination lasts from the earliest times until
temperatures of order a few eV (when the universe is
of order $10^5$ years old or so).  During the radiation
dominated era, the energy and entropy densities are:
$$\rho= {\pi^2 \over 30} g T^4~~~~~s
= {2 \over 45} \pi^2 g T^3~~~~~g=g_B + 7/8 g_F. \eqn\rhoands$$
During this period, $t \sim T^{-2}$.
We can now list a few highlights in the history of the universe:
\item{1.}  $T \approx 2.7 K$:  the present moment ($t \approx
13 \times 10^9$~yrs)
\item{2.}  $T\sim 10^o K$ ($t \sim 10^9$ yrs):
earliest formation of galaxies.
\item{3.}  $T \approx$ few eV ($t \sim
10^4$ yrs):  transition from radiation
to matter domination
\item{3.}  $T \approx  ~0.4$ eV ($t \sim 10^5$ yrs)
recombination
\item{4.}  $T \approx$ MeV  ($t\sim 1$ sec):
this is the moment
of neutrino decoupling.  At this point, weak interactions no longer
maintain the equilibrium between protons and neutrons.
Subsequently, at $T \approx 0.1$ MeV  ($t \approx 3$ minutes)
those neutrons which
have not yet decayed are bound into the light nuclei $^4$He, $^2$D,
$^3$He, $^7$Li.  One of the great triumphs of the standard
cosmology is its successful prediction (through detailed
calculations) of the abundances
of these ``primordial" elements.  This success, however, only
holds if the density of baryons lies in a narrow range:\refmark{
\nucleosynthesis}
$$0.011 < \Omega_b h^2 <
0.048~~~~~N_{\nu}<5\eqn\nucleosynthesis$$
where $\Omega_b$ is the fraction of the critical density in baryons.
\item{5.}  $T \approx 200$ MeV:  The QCD phase transition -- QCD
passes from an unconfined, quark-gluon plasma phase to a
confined phase (gas of nucleons and pions).
\item{6.}  $T \approx 100~\gev\ $:  The electroweak phase transition.
Above this temperature, the $W$ and $Z$ bosons are essentially
massless, and the Higgs particles (or whatever) have no vacuum
expectation values.  Perhaps the baryon asymmetry,
$n_b/n_{\gamma} \sim 10^{-10}$ is created at this temperature.
\item{7.}  Still higher temperatures:  Here we enter into a still more
speculative realm.  If a Peccei-Quinn symmetry exists, there
is a phase transition associated with this, presumably (but not
necessarily) at a temperature of order $f_a$, the axion decay constant.
Many other phenomena may occur as well.
\item{8.}  At some very high temperature (perhaps as low as the
weak scale, but probably higher\refmark{\knox}),
inflation occurs.

\REF\weinbergetal{Y. Hu, M.S. Turner and E.J. Weinberg,
CU-TP-581 (1993).}
Inflation is not really the subject of this talk, but let me say a few
words about it here.  Inflation, at the present time, is a generic
term, referring to a class of phenomena which could solve a number
of serious puzzles with big bang cosmology.
(Indeed, it has recently been argued that inflation is the
unique solution to these problems.)\refmark{\weinbergetal}
Briefly, these problems are:
\item{1.} The  universe is remarkably
homogeneous and isotropic on very
large scales.  From COBE, we know that the temperature of the
microwave background exhibits
only tiny variations on angular scales of $10$ degrees or so.
If we just run the big bang clock backwards, this means that
about $10^5$ regions which were causally disconnected at
recombination have nearly the same temperature.   This would
seem to violate causality.
A similar statement applies to the synthesis of the light
elements (at the time of nucleosynthesis, our
present universe corresponds to about $10^{25}$ causally
disconnected regions).
\item{2.}  The $\Omega$ problem:  $\Omega$ behaves with time as
$$\Omega= {1 \over 1-x(t)}~~~~~~
x= {k \over a^2} {1 \over 8 \pi G \rho/3}\eqn\omegat$$
where $a$ here is normalized so as to be $1$ today.  But since in the
far past, $a$ was many orders of magnitude smaller (\eg,
$10$ orders of magnitude at nucleosynthesis),  and given that
$\Omega$ is within a few orders of magnitude of unity today,
it was {\it extremely} close to unity at early times.
\item{3.}  Particle physics models often predict stable, heavy particles,
such as mono- poles, and topological objects, such as domain walls.
These would yield far more mass than is currently observed.

In inflationary schemes, there is
a period in which the energy density of the universe is dominated neither
by radiation nor by matter, but by a cosmological constant -- essentially
vacuum energy.  During such a period, Einstein's equations yield
that
$a \sim e^{Ht}$
where
$H^2 = {8 \pi G V/3 V}$
A simple model for such a process is provided by the ``new inflationary
scenario"
(by now a misnomer).
In this scenario, there is a scalar field,
$\phi$ (the ``inflaton").  By assumption, the potential of this field is
extremely
flat.  At early times, the inflaton does not sit at the minimum of its
potential, due, for example, to thermal effects.  Instead, it sits near the
origin.
At some time, it begins to roll, very slowly, towards the minimum.
During this period, there is, effectively, a non-zero cosmological constant.
If the period of exponential growth lasts for a sufficient number of e-foldings
(\eg, $100$'s),
all of the problems mentioned above are solved.  First, small, causally
connected
regions of the universe grow to enormous size -- sizes larger than our
observable
universe.  This solves the homogeneity and flatness problems.  Second, if such
a region contains, say a single monopole, then there is at most one monopole
in our observable universe; more generally, any stable objects will be diluted
away.  Third, provided the universe reheats significantly
after inflation, a great deal of entropy is created.  This occurs, for example,
if
the inflaton is reasonably strongly coupled at the end of inflation.  In this
case,
as it sloshes around near the minimum, its motion damps, and the energy
is dissipated into heating of the ambient plasma.  A vast amount of entropy can
be
created in this way.

There are many problems with this particular scenario, and many variants
are currently on the market.  I don't want to review these here, but simply
stress that there are two predictions generic to these schemes:
\item{1.}  $\Omega=1$.  (See eq.~\friedman; essentially,
the $k$ term becomes irrelevant due to the large value of $a$.)
\item{2.}  The description I gave of the inflationary model above was
completely
classical.  Quantum effects lead to fluctuations (\eg, in the position of the
field,
$\phi$, in time) and these lead to fluctuations in the density as a function of
wavenumber, $k$.  The wavelength
is ``stretched" by expansion; once these fluctuations are larger
than the horizon size (essentially $H^{-1}$), they are
frozen.  In this way one obtains an essentially scale-invariant
fluctuation of density fluctuations (``Harrison-Zeldovich spectrum").
Once inflation is over, the universe continues to expand.
As a result, at any given time, fluctuations on a scale of order
the horizon size reenter the horizon.  When they do, they
can begin to grow.  However, this growth is only significant
during the matter dominated
era, when  ${\delta \rho / \rho} \propto a(t)$.
Once ${\delta \rho / \rho} \sim 1$,
the system becomes non-linear, and structure begins to
form.  (The reader should be aware that there exist competing,
so-called ``non-Gaussian" models of structure formation, such
as textures and cosmic strings.)

\FIG\cobefig{Correlation functions observed by COBE compared
with the predictions of a Harrison-Zeldovich spectrum.}
Where does COBE fit into this?  COBE found fluctuations in
the sky in the microwave temperature of approximately a part
in $10^{-5}$.
The fluctuations in the energy
density are related to fluctuations in the temperature by
an equation of the form
$${\delta T \over T} \approx {1 \over 10} {\delta \rho \over \rho}
.\eqn\sachswolfe$$
As we will see when we discuss dark matter,
the COBE result is consistent (to at least a factor of two)
with a picture in which galaxies begin to form when
these fluctuations became non-linear.  Moreover, COBE observed
a power spectrum, $P(k) = A k^{1.1 \pm 0.5}$, consistent
with Harrison-Zeldovich.  (See fig.~\cobefig.)

\midinsert
   \tenpoint \baselineskip=12pt   \narrower
\vskip9.5cm
\vskip12pt\noindent
{\bf Fig.~\cobefig.}\enskip
{Correlation functions observed by COBE compared
with the predictions of a Harrison-Zeldovich spectrum.}
\endinsert

\chapter{DARK MATTER}

\REF\peculiar{M. Rowan-Robinson \etal, Mon. Not. R. Astr.
Soc. {\bf 247} (1990) 1; N. Kaiser \etal, {\it ibid.} {\bf 282} (1991) 1;
M. Strauss \etal, Astrophys. J. {\bf 385} (1992) 444;
E. Bertschinger and A. Dekel, Astrophys. J. {\bf 336} (1989) L5;
A. Dekel \etal, Astrophys. J., in press; M. Strauss \etal,
{\it ibid.} {\bf 397} (1992) 395.}
Inflation, as well as general fine-tuning arguments strongly suggest that
$\Omega=1$.  What is the observational situation?
There is only enough luminous matter (stars
we can see) to give $\Omega h^2 = 0.005$--$0.007$.
 But it has been known for a long time
that there is substantially more matter which we can't see -- the
infamous dark matter.
The most direct evidence of this comes
from examining rotation curves
of spiral galaxies (of which our own Milky Way is an example),
\ie, how the velocities of stars vary with distance from the center.  Just
using Newton's laws, one finds that the force on the outlying stars is
much too large to be accounted for by the visible matter.  Indeed, the
galaxy has a halo, usually assumed to be spherical,
with a density which falls more slowly than the visible density.
 From general studies of such rotation curves, astronomers estimate
that this missing matter contributes to $\Omega$ an amount between
$0.0 3$ and $0.1$.  On larger scales,
there is evidence for even larger densities of dark matter.
(The issue, here, is how the dark matter has ``clumped.")  From
studies of groups and clusters of galaxies, one obtains $\Omega
=0.05$--$0.3$.  Perhaps most interesting are recent studies of
peculiar velocities (IRAS and POTENT)  One studies
the motion of galaxies (particularly ours)
relative to the CMBR (about 620 km/sec), and assumes that this
can be understood as arising from the galaxies in a large galaxy survey.
These studies give
$\Omega/b \sim 1.2 \pm 0.6$.\refmark{\peculiar}  Here $b$ is the so-called
biasing factor; it is related to how luminous mass traces the dark
matter.

So it is well established that most of the matter in the universe is
non-luminous,
and it may well be true that $\Omega=1$.  Recalling eq.~%
\nucleosynthesis,
we see that big bang nucleosynthesis allows a
baryon number density consistent
with the missing mass in the galactic halo, but not with $\Omega=1$.
For the rest of this talk, I will adopt the working assumption that there
is non-baryonic missing matter.  Even then, however, there is the
question
of whether the matter in the galactic halo is baryonic (perhaps the
non-baryonic
matter clumps only on scales larger than galaxy scales).

\REF\greist{For a brief overview, with refs., see
K. Griest, UC San Diego preprint
UCSD-ASTRO-TH-92-01.}
\REF\nytmacho{The New York Times, Sept. 22, 1993, B7.}
This  question of whether the galactic halo contains baryonic matter is
subject to experimental test.  In particular, the baryonic matter might
be in the form of MACHO's, ``massive compact halo objects." These could be
``jupiters," objects with mass
of order $0.001 M_\sun$, or brown dwarfs,
with mass $0.01 M_\sun$.\refmark{\greist}
In either case, these objects,
consisting principally of hydrogen and helium,
would be too light to ignite nuclear burning.
There are already several searches for such objects underway.  The idea
is to look for ``microlensing," multiple images (actually enhanced images,
since the multiple image, or ring, cannot be resolved on earth)
which result when one of these objects passes between our line of sight
and a star.  Because the distribution of non-luminous matter in the
galaxy is known, one can predict the expected rate.
There are actually three searches underway,
looking for MACHO's in our own galactic
halo by looking for intensity variations of stars
in the Large Magellenic Cloud.  The
typical time scale for these variations are of order one
week (one week for 0.1 $M_\sun$ objects;
the time varies as $\sqrt{M}$).
The MACHO collaboration, for example,
(LLNL, CPA, Mt.~Stromlo Observatory) has exclusive use of a 1.3 m telescope
for at least 4 years.  They observe about 10--20 million stars per
night.  They would expect
something of order $500$ events
per year for a halo density of jupiter mass objects
(0.001 M) (with events lasting about 30 days).
So far, they have analyzed
one data set, containing 1.7 million stars, observed for about
nine months.   In this sample, they would have expected to
see a few brown dwarfs; with their cuts, however, they
have no candidates.  If the cuts are relaxed, there are a handful
of candidates.  These they believe are probably variable stars;
if they are real microlensing events, there will be many candidates
in the next set.
Within about four years time, definitive results can
probably be obtained for
objects with masses between $10^{-5}$--$10^2~M_\sun$.
This is, in fact, the entire interesting range; smaller
objects would have already evaporated, while larger
ones would disrupt the galactic disk.
So it is possible that within
a few years we will know with some certainty whether the dark matter in
the galaxy is baryonic or not.  Indeed, as this document was
``going to press," there were preliminary announcements of
macho candidates.\refmark{\nytmacho}

COBE provides indirect evidence that the dark matter is not baryonic.
In a baryon-dominated universe,
fluctuations which enter the horizon prior
to decoupling are washed out.  Since decoupling, the
scale factor has grown by about $10^3$, so fluctuations
as small as $10^{-5}$ would not even now be non-linear.
An $\Omega=1$ universe, dominated by non-baryonic matter
does much better, because the fluctuations start to grow
upon matter domination, and matter domination occurs earlier.

\REF\krauss{For a discussion of possible deviations from
a scale invariant spectrum, see, for example,
L. Krauss, Case preprint CASE-P1-93 (1993).}
Assuming that most of the matter in the universe is non-baryonic, one usually
distinguishes two types of dark matter.  Cold dark matter
is defined as matter which is relativistic when it
drops out of equilibrium (``freeze out"),  while hot dark
matter is relativistic at freeze out.
In the case of cold dark
matter,  this leads  to a quite compelling theory of structure on
the scale of galaxies and clusters of galaxies.
Hot dark matter
does not appear to clump enough
on galactic scales, particularly of small dwarf galaxies.
However, cold dark matter has trouble with the recent
COBE results, if one makes the standard assumption---motivated
by the simplest inflationary models---of a
scale-invariant spectrum (there is too little
power at large scales, too much at small if CDM normalized
to galaxy-galaxy two-point function).
To fix this, there have been various
proposals.  These include modification of the fluctuation
spectrum.
This does not seem unreasonable, given that most
present theories of
inflation are not completely satisfactory.  For example, if the potential
responsible for inflation changes during the
inflationary epoch, this can lead to departures from scale
invariance.\refmark{\krauss}
Another possibility is that the ``bias," which we
have discussed earlier, is not constant with scale.
Perhaps the most
popular alternative at the moment, however, is to suppose that there
is actually a mixture of cold (70\%) and hot (30\%)
dark matter, where the hot dark matter might be a neutrino
with mass of order 7 eV.\refmark{\hcdm}
Of course, this requires a quite remarkable coincidence,
and I will leave it to you to decide how plausible such a coincidence
might be.  Still, 7~eV is not an unreasonable value for a neutrino
mass (\eg, see-saw mechanisms), and this model has
the virtue that it fits quite a range of data
with only one extra parameter.
Since I am not going to focus heavily on neutrinos in this talk,
for the rest I will simply adopt the viewpoint that dark matter certainly
exists, and there is a good chance that some of it is in an exotic
form.  Note that even in this cold plus hot scenario, on galactic
scales, the dark matter is principally cold, so experiments
designed to look for the dark matter in the halo are unaffected.

For the moment, let us not worry about the details of primordial
fluctuations and the development of structure, but rather
consider some of the dark matter candidates suggested by
particle physics.
The two which will interest us here are WIMP's (Weakly Interacting
Massive Particles) and axions.  As we will see, massive particles
with cross sections of weak interaction magnitude are ideal
dark matter candidates.  It is very easy to cook up models with
such particles, but only a small number are very well motivated.
Of these, Majorana and Dirac neutrinos are ruled out by
LEP and direct dark matter searches.  This leaves the neutralino,
a particle suggested by supersymmetry, as the most promising
of the (currently) well-motivated candidates.  This candidate is so
plausible that over the last year or so model builders have been
using the condition $\Omega=1$ to constrain the supersymmetry
parameters.  Meanwhile, a variety of terrestrial searches have a
very real prospect of seeing this dark matter, if it exists.

The second well-motivated candidate for dark matter is the axion.
Its role as dark matter arises in a quite different way from that of WIMP's,
and if it is found
it will provide a window on physics at much higher energies than we can
contemplate exploring with accelerators -- $10^{12}~\gev$ or so.
We will turn to this particle first.

\chapter{AXIONS}

\REF\muzero{H. Georgi and I. McArthur, Harvard
University Report No. HUTP-81/A011 (unpublished);
D.B. Kaplan and A.V. Manohar,  Phys. Rev. Lett. {\bf
56}, 2004 (1986); K. Choi, C.W. Kim and W.K. Sze,  Phys. Rev. Lett.
{\bf 61}, 794 (1988); J. Donoghue and D. Wyler, Phys. Rev. {\bf D45}
(1992) 892; K. Choi, Nucl. Phys. {\bf B383} (1992) 58.}
\REF\nirseiberg{Y. Nir and N. Seiberg, Nucl. Phys. {\bf B398} (1993)
319.}
\REF\nelson{A. Nelson, Phys. Lett. {\bf 136B} (1984) 387.}
\REF\barr{S.M. Barr, Phys. Rev. Lett. {\bf 53} (1984) 329.}
\REF\dkl{M. Dine, A. Kagan and R. Leigh, SCIPP 93/05 (1993).}
\REF\axionreviews{For reviews, see, for example, G.G.
Raffelt, Phys. Rep. {\bf 198} (1990) 1; M.S. Turner, {\it ibid.}
{\bf 197} (1990) 67.}
\REF\sikiviereview{P. Sikivie, University of Florida preprint
UFIFT-HEP-92-25.}
QCD is a quite
successful theory of strong interactions.  However,  in addition to
the well-known $\Lambda$ parameter, this theory has another
parameter:  it is possible to add to the lagrangian a term
$${\theta g^2 \over 32 \pi^2} F_{\mu \nu}^a \tilde
F_{\mu \nu}^a.\eqn\thetaparameter$$
This term is $CP$-violating.  Formally, it is a total divergence,
but arguments based on current algebra can be used to
show that it has a definite effect on physics:  it leads to
too large a value of the electric dipole moment unless
$\theta<10^{-9}$.  One can view this as just one more small
coupling (like the electron Yukawa coupling) which we
don't know how to explain.  But in fact there are three proposals
for understanding the small value of this number:
\item{1.}  $m_u=0$.  This contradicts the
usual, first order current algebra analysis of the pseudoscalar
masses, which gives $m_u/m_d \approx 0.55$.  But it has been
argued (rather convincingly in my view) that second order
corrections to this relation are large, and that one cannot
rule out $m_u=0$.\refmark{\muzero}
It is also often argued that  it is unnatural
to have $m_u=0$.  Any symmetry which might protect the
$u$ quark mass would necessarily be anomalous (in a moment,
however, we will see that
a similar argument applies -- in spades -- to axions), and
rather puzzling from the perspective of, say, grand unification.
It has recently been pointed out, however, that
such anomalous discrete symmetries do arise in string theory.
Recently proposed schemes for understanding quark mass
matrices also frequently lead to $m_u=0$.\refmark{\nirseiberg}
Thus, while our focus here will be on axions, one should keep
in mind that this  first solution of the strong $CP$ problem
is quite plausible.
\item{2.}  $CP$ is a good symmetry of nature,
so there is no ``bare" $\theta$, and is spontaneously
broken in such a way that the effective $\theta$ is small.
Some time ago, Nelson and Barr suggested a mechanism
which would give a non-zero KM phase, while at the same
time giving a very small $\theta$.\refmark{\nelson,
\barr}  Unfortunately, in
the framework of supersymmetry, it has recently
been shown that loop corrections to $\theta$ are generically
quite large in such schemes.\refmark{\dkl}
\item{3.}  Axions.\refmark{\axionreviews,\sikiviereview}
The basic idea for solving the axion problem, due to Peccei and
Quinn, is to make $\theta$ a dynamical variable, by introducing
a field $a(x)$, which couples to $F \tilde F$:
$$\left[Na(x)/f_a +\theta\right]{g^2 \over 32 \pi^2} F_{\mu \nu}^a \tilde
F_{\mu \nu}^a.\eqn\axioncoupling$$
$f_a$ is called the axion decay constant.
The rest of the lagrangian is assumed to be symmetric under the
shift (``Peccei-Quinn symmetry")
$${a \over f_a}\rightarrow {a \over f_a} + \delta.
\eqn\axiontransformation$$
By a symmetry transformation we can
remove $\theta$; the effective $\theta$ is then just the
expectation value of $a$.  In QCD, it is easy to show that
this vev is very tiny, because,
in the absence of $\theta$ the theory conserves $CP$.

To actually construct models of this phenomenon, one typically
considers theories with a scalar field, $\phi$, which transforms
under an anomalous $U(1)$ symmetry.  $\phi$ is assumed to
obtain a vev, $v=f_a$.  One then writes
$$\phi={v \over \sqrt{2}}\ e^{i a(x)/v}.\eqn\phiparameterized$$
The coupling to $F \tilde F$ then arises through then anomaly.
QCD effects give rise to a potential for the axion, which can
be computed using ordinary current algebra.  One finds
$$m_a = \vert N \vert {f_{\pi} m_{\pi}\over f_a}
{\sqrt{ m_u m_d} \over m_u + m_d} = 0.6\ \ev\
{10^7\ \gev\ \over f_a/\vert N \vert}\eqn\axionmass$$
where
$N= 2 \sum_f t_f Q_f^{PQ}$,
the sum is over fermion species, and $t_f$ is an appropriate Casimir.
There is also, typically, a coupling to two photons,
$${\cal L}_{a \gamma \gamma} = {\alpha \over 8 \pi}
{a \over f_a} \left[ N_e/N - \left(5/3 +
{m_d -m_u \over m_d + m_u}\right) \right] F \tilde F.\eqn\axionphoton$$
$N_e$ reflects the QED anomaly of the Peccei-Quinn symmetry;
$8/3$ is a value which typifies a broad class of models.
With this choice,
$${\cal L}_{a \gamma \gamma}\approx {\alpha \over \pi}
 {m_a \over 0.6 \times 10^{16}~\ev^2} a(x) \vec E \cdot \vec B.
\eqn\axionphotonnumerical$$
One can also work out couplings of axions to fermions.

Note that if $f_a=M_p$, the axion is extremely weakly coupled,
but also extremely light, so assuming initially a
thermal distribution of axions, their density today
would be comparable
to that of photons.  Why, then, are axions a plausible dark matter
candidate?  Suppose that inflation occurs after the PQ phase transition, \ie,
at
temperatures such that $\phi$ has a non-zero
vev.  At extremely high temperatures,
the QCD effects which give rise to the axion potential are so
small (due to asymptotic freedom) that $\VEV a$ is essentially a random
variable.  In particular,
prior to inflation, regions of the order of the horizon size
have different values of $\VEV a$ ($\theta$).  After inflation,
the region which will become our present universe
has a single value of $\VEV a$.
As the universe cools towards QCD temperatures, the
axion potential ``turns on."  The axion field will then roll towards
its minimum.  However, because of its weak coupling, it cannot
dissipate energy effectively, and simply oscillates.  Indeed, the
axion field carries an energy of order
$$V = m_a^2 a^2
\sim (100~\mev)^4~{a^2 \over f_a^2}.\eqn\roughpotential$$
One can think of this
field as a coherent state of axions, where the density of axions
is $|V|/ m_a$, a huge number!
This axion density dilutes like matter (it falls as $T^3$, whereas
the radiation falls as $T^4$), so the axions eventually
come to dominate the energy density.  Putting in the numbers,
one finds
$$\rho_a = \rho_{crit}\theta_o^2
\left({0.6 \times 10^{-5}~\ev \over m_a}\right)
\left({200~\mev \over \Lambda_{QCD}}\right)^{3/4}
\left({75~{\rm km\ s}^{-1} {\rm Mpc}^{-1} \over H_o}\right)^2
\eqn\axiondensity$$
where $\theta_o={a_o / f_a}$ is the initial value
of the effective $\theta$ angle.
One sees that axions overclose the universe unless
the axion mass is not too small, or we happen
to be living in a part of the universe where $\theta_o$ happens
to be very small.  For a range of axion masses around
$10^{-5}$~eV, the axions are a dark matter candidate.
They are cold dark matter because they are highly non-relativistic.

\REF\sikiviewalls{P. Sikivie, Phys. Rev. Lett. {\bf 48} (1982) 1156.}
\REF\davis{R. Davis, Phys. Rev. {\bf D32} (1985) 3172; Phys. Lett.
{\bf B180} (1986); R. L. Davis and E.P.S. Shellard, Nucl. Phys.
{\bf B324} (1989) 167}
\REF\sikiviestrings{D. Harari and P. Sikivie, Phys. Lett. {\bf 195 B}
(1987) 361; C. Hagmann and P. Sikivie, Nucl Phys. {\bf B363} (1991)
247.}
\REF\lythe{D. Lythe and E. D. Stewart,
Phys. Rev. {\bf D46} (1992) 532.}
$m_a > 10^{-5}$ corresponds to $f_a < 10^{12}$ GeV, so
it is possible that the
Peccei-Quinn transition occurs {\it after} inflation.
If this is the case, the situation is distinctly more complicated.
First, there is the danger of forming domain walls.\refmark{\sikiviewalls}
The point is that, in general, QCD does not break the
Peccei-Quinn symmetry completely, but typically leaves
over a discrete $Z_N$ symmetry.  I know of two solutions
to this problem.  First, there is the possibility that $N$ is simply
equal to one.  Second, as Sikivie pointed out long ago,\refmark{
\sikiviewalls} if one
had small effects which explicitly break the Peccei-Quinn symmetry,
these could lead to collapse of the domain walls.  As we will
discuss shortly, the Peccei-Quinn symmetry, if it exists, is likely
to be an accidental, approximate symmetry.  It is almost certainly
violated by higher dimension operators, so the symmetry-violating
terms envisioned by Sikivie are quite plausible.  Because the
Hubble constant at this time is about $19$ orders of magnitude
smaller than the QCD scale, quite small corrections can cause
the collapse of the walls on a cosmologically short time scale.

As stressed by Davis,\refmark{\davis} and discussed
by Sikivie and collaborators\refmark{\sikiviestrings} at the,
Peccei-Quinn phase transition, one expects to
form networks of axion strings, which in turn are a coherent
source of axions.  There seems to be some controversy in the
literature about just how large an axion density one makes, with
estimates ranging from values similar to those of eq.~\axiondensity\
to values $100$ times a large (which might rule out axions all together).
For the rest of this discussion, we will simply assume inflation occurs first,
but these other possibilities should be kept in mind.\foot{Recently,
Lythe\refmark{\lythe} has pointed out that if during inflation, $H \sim f_a$,
there can still be problems with strings.}

\FIG\sikiviefig{Figure from ref.~\sikiviereview, indicating
allowed range of axion masses and decay constants.}
\REF\raffeltseckel{G. Raffelt and D. Seckel, Phys. Rev.
Lett. {\bf 67} (1991) 2606.}
\REF\sklein{S. Klein, these proceedings.}
There is an upper bound on the axion mass, which comes from
more conventional astrophysics.  The axion mass is
proportional to the strength of the axion interaction
(inversely proportional to $f_a$).  As a result, if axions are too
heavy, they are copiously produced in stars.  However, they
still interact sufficiently weakly that they escape the star, carrying
off energy.  The strongest bounds of this type come from
supernova SN 1987a. (It should be noted, however, that
these bounds are probably weakened somewhat
by the ``LPM effect," as pointed out by Raffelt and
Seckel.\refmark{\raffeltseckel}
The first experimental observation
of this effect was described in S. Klein's talk at this
meeting.\refmark{\sklein})
This combination of bounds is indicated
in fig.~\sikiviefig.~ So if axions exist at all, and the
cosmological arguments are correct, they are in the right range
to  be a dark matter candidate.

\midinsert
   \tenpoint \baselineskip=12pt   \narrower
\vskip13.5cm
\vskip6pt\noindent
{\bf Fig.~\sikiviefig.}\enskip
{Figure from ref.~\sikiviereview, indicating
allowed range of axion masses and decay constants.}
\vskip6pt
\endinsert

\REF\sikiviedetector{P. Sikivie, Phys. Rev. Lett.
{\bf 51} (1983) 1415.}
Can axions be detected
if they constitute the halo of our galaxy?  Pierre Sikivie has
for some years been advocating searches for such
axions,\refmark{\sikiviedetector}
and by now there have been at least two prototype experiments,
and a full scale experiment which can study an interesting
range of parameters has been proposed and approved.   The idea
is to stimulate axion conversions to photons in a cavity placed
in
a strong magnetic field, using the coupling in ${\cal L}_{a \gamma
\gamma}$.
The axions can then, in the presence of the field, excite
a mode of the cavity (a $TM_{nlo}$ mode, since
the size of the cavity is typically much less than the de Broglie
wavelength of the axion, $\lambda_a \sim 2 \pi 10^3 m_a^{-1}$,
where the first factor reflects the typical velocity of axions
in the halo).

\FIG\prototype{Results of prototype axion search experiments
showing the parameter range ruled out.  This is taken
from the proposal described in the text.  Range which may
be excluded by future experiments is indicated.}
One can then compute the power in this mode; it behaves as
$$\eqalign{%
P_{nl} &= 2 \times 10^{-26}~{\rm watt}~
\left({V \over 500~{\rm liter}}\right)
\left({B_o \over 8~ {\rm Tesla}}\right)^2
C_{nl} {\rho_a \over 10^{-24}~{\rm gm/cm}^3}
{m_a \over 2 \pi \times 3~{\rm GHz}}   \cr
&~~~\times \min[Q_L, Q_a] \cr}
\eqn\axionpower$$
where $Q_L$ and $Q_a$ are respectively the quality
factors of the cavity and the galactic signal (the ratio
of energy to energy spread), and $C_{nl}$ is a geometric
factor.  Part of the difficulty of these experiments comes
from the fact that one must study a large range of frequencies
in very small intervals.  The cavity must thus be tunable.
Prototype experiments have already been carried out at
BNL (Rochester, Brookhaven, Fermilab) and at Florida.
The results are shown in fig.~\prototype, where they are compared
with the predictions of two popular axion models.
Note that the sensitivity of these experiments is about
two orders of magnitude too low to rule out (or establish!)
these two models.
These experiments both had $B^2 V$ about $0.4 T^2 m^3$.
The Florida experiment had somewhat greater sensitivity due
to more efficient data taking and better microwave
equipment.

\midinsert
   \tenpoint \baselineskip=12pt   \narrower
\vskip7cm
\vskip6pt\noindent
{\bf Fig.~\prototype.}\enskip
{Results of prototype experiments
showing the parameter range ruled out.  This is taken
from the proposal described in the text.  Range which may
be excluded by future experiments is indicated.}
\endinsert

\REF\highmass{See, \eg, P. Sikivie, D.B. Tanner and Y. Wang,
University of Florida preprint UFIFT-HEP-93-2 (1993).}
There is now an approved proposal for an experiment using
a new, much larger magnet, with
$B=8.5 T$ and a 60 cm diameter (permitting
installation of a 50 cm cavity).
(The original proposal involved a decomissioned
magnet from a mirror fusion test facility at LLNL.)
Delivery of the magnet is expected in April of '94.  Eventually,
with multiple cavities, it should be possible to search
to a mass of $12.6\ \mu$eV, and possibly higher.  There
is also the possibility of installing cavities in a $14T$
magnet at the National Magnet Lab.  However, this magnet
is only available for a brief period, so there
is discussion of building a dedicated magnet of this size, as well.
Other suggestions have been made for observing axions of larger
mass.\refmark{\highmass} These proposals involve cavity detectors,
but also involve various alternative detectors, \eg, involving
superconducting wires.  Current in the wires gives a
spatially varying magnetic field, which is used to enhance
axion production.

Having described a little bit these wonderful experiments,
let me step back and ask a rather embarrassing question,
which has been raised by a number of authors recently:
just how plausible {\it is} the axion solution of the strong
$CP$ problem?  The question is not really a new one.
After all, the Peccei-Quinn symmetry is a puzzling one:
a symmetry which is not really a symmetry.  Moreover, as
we have already mentioned,
the attitude has developed among theorists in recent years
that there should be no global symmetries in nature.
More precisely, any global symmetries which exist should
be accidents of renormalizability, just as baryon number
is an accidental symmetry of the standard model.  What the
recent discussions have made clear is that if the
Peccei-Quinn symmetry
is an accident of this type, the accident must be an
extraordinarily good one; operators of very high dimension must be
suppressed.

The difficulty is easy to understand;
\REF\wise{H. Georgi, S. Glashow and M. Wise, Phys. Rev. Lett.
{\bf 47} (1981) 402.}
\REF\march{M. Kamionkowski and J. March-Russell, Phys. Lett.
{\bf 282B} (1992) 137; R. Holman {\it \etal},
Phys. Lett. {\bf 282B} (1992) 132.}
it was noted in passing by Georgi, Glashow
and Wise.\refmark{\wise}
More recently, it has been discussed
in a general and quantitative
fashion by several authors.\refmark{\march}
To gain some appreciation of the difficulty,
suppose that the lowest dimension,
 gauge-invariant operator which violates the symmetry
is ${\cal O}^{(4+n)}$, of dimension
$4+n$.  Then the leading symmetry-violating term which can occur in
a low-energy effective field theory is
$${\cal L}_{SB} = {\gamma \over M_P^n} {\cal O}^{(4+n)}\eqn\lsb$$
where $\gamma$ is a dimensionless coupling constant.
On dimensional grounds, this gives rise to a linear term in the
axion potential,
$$V_{SB} \propto \gamma{f_a^{n+3} \over M_P^n} a(x).\eqn
\linearterm$$
Since
$m_a^2 \sim {m_{\pi}^2 f_{\pi}^2 / f_a^2}$
the resulting shift in $\theta$ is
$$\delta \theta = {\delta a \over f_a} \sim {\gamma \over m_{\pi^2}
f_{\pi}^2} {f_a^{n+4} \over M_P^n}~<~ 10^{-9}.\eqn\thetashift$$
For $f_a = 10^{11}$, this gives $n>7$ (\ie, the symmetry-violating
operator must at least be of dimension $12$!)  If $f_a = 10^{10}$,
things are slightly better; one needs to suppress all operators
of dimension less than $9$.  Of course, if $f_a$ is larger,
one must forbid an even larger number of operators.

\REF\wittenaxion{E. Witten, Phys. Lett. {\bf B149} (1984) 359.}
\REF\anthropic{A. Linde, Phys. Lett. {\bf B201} (1988) 437;
Phys. Lett. {\bf B259} (1991) 38.}
This all sounds rather hopeless.  But in string theory,
it has long been known that, in perturbation theory,
the Peccei-Quinn symmetry is exact!\refmark{\wittenaxion}
One way to understand
this is as an accidental consequence of another gauge
symmetry of the theory, involving the antisymmetric
tensor field.  Unfortunately for our present considerations,
this axion has $f_a \sim M_p$.  Only if we give up the
cosmological bound is such an axion acceptable.
One can imagine a number of ways in which this bound
might be relaxed.  For example, there might be some
generation of entropy after the  QCD phase transition.
Or perhaps, as a result of some sort of anthropic
considerations, the initial value of the $\theta$-angle
in our observable universe is small, of order $10^{-3}$ or
less (the axion energy density is proportional to the square
of this angle, though we have not indicated it explicitly
above).  One proposal for
such an anthropic explanation has been made
in ref. \anthropic.  However, such an axion would be too weakly coupled
to be detectable in any of these experiments, even if it
did make up the dark matter of the halo.

If we do take the cosmological bound seriously,
the lesson of all this is that if one wants a Peccei-Quinn
symmetry to arise by accident, one must forbid operators
up to very high dimensions.  How might such a thing occur?
The authors of refs. \march\ noted that
with a sufficiently complicated continuous gauge symmetry,
one could indeed suppress operators of very high dimension.
However, by their own admission, the resulting models were not
particularly beautiful.

\REF\shafi{G. Lazarides, C. Panagiotakopoulos and Q. Shafi,
Phys. Rev. Lett. {\bf 56} (1986) 432.}
\REF\ross{J. Casas and G. Ross, Phys. Lett. {\bf 192B} (1987) 119.}
\REF\cinti{M. Dine, in {\it Topics in Quantum Gravity and Beyond},
F. Mansouori and J. Scanio, eds., World Scientific,
Singapore (1993).}
In my view, a more plausible explanation for an axion with
a decay constant of order $10^{12}$ or so is as an accidental
consequence of a discrete symmetry.
In fact, in the framework of string theory, such a possibility
was considered long ago by
Lazarides \etal\refmark{\shafi}
and by Ross and Casas.\refmark{\ross}  The latter authors also attempted
to estimate how large a $\theta$ would be induced by
higher-dimension operators which violated the Peccei-Quinn
symmetry, in precisely the spirit described above (it turns out
that they neglected an important class of operators, but
this difficulty is easily remedied\refmark{\cinti}).
Rather than review these models in detail, however, it
is useful to illustrate just how
powerful discrete symmetries
are in this respect by considering theories in
which the Peccei-Quinn symmetry is dynamically broken by fermion
condensates.\foot{This has been noted
independently, and much earlier, by A. Nelson (unpublished).}
As an example, consider a theory with (unbroken)
gauge group (in addition to the standard
model gauge group) $SU(4)_{AC}$ ($AC$ is for ``axi-color"), with
scale $\Lambda_{AC} \sim f_a$.  In addition to the usual quarks
and leptons, we suppose that the theory contains additional fields
$Q$ and $\bar Q$, transforming as $(4,3)$ and $(\bar 4, \bar 3)$ under
$SU(4)_{AC} \times SU(3)_c$, and fields ${\cal Q}$ and
$\bar {\cal Q}$ transforming as a $(4, 1)$ and a $(\bar 4, 1)$.
Now suppose that the model possesses a discrete symmetry (gauged
or global) under which
$$Q \rightarrow \alpha Q~~~~~~~~{\cal Q} \rightarrow \alpha {\cal
Q}\eqn\transforma$$
where $\alpha = e^{2 \pi i / N}$; all other fields are neutral.
If, for example,
$N=3$, the lowest dimension chirality-violating operators
one can write are of the form $(\bar Q Q)^3$, which is dimension $9$;
suppression of still higher dimension operators is achieved by choosing
larger $N$.  In this theory, the would-be PQ symmetry is
$$Q \rightarrow e^{i \omega Q}~~~~~~~~{\cal Q} \rightarrow\
e^{-3 i \omega} {\cal Q} .\eqn\pqtrans$$
This symmetry has no $SU(4)$ anomaly, but it does have a QCD anomaly.
One expects that this symmetry will be broken by the
condensates
$$\VEV{\bar QQ}\sim \VEV{\bar{\cal Q}{\cal Q}} \sim f_a^3.\eqn\condensates$$
This gives rise to an axion with decay constant $f_a$, which solves
the strong $CP$ problem.


Lazarides \etal\ and Casas and
Ross wrote down string inspired models which accomplished the
same objective as in the model above.  Again, discrete symmetries
suppressed operators up to very high dimension.
These models have a major virtue:  the axion decay constant is naturally
of order $M_{INT}=\sqrt{M_{W} M_P}$, \ie,
within the allowed axion window.

If, for the moment, we accept this picture as the origin of the
Peccei-Quinn symmetry, there is one interesting cosmological
consequence, which we have already alluded to.  This is the
point that the symmetry is not likely to be exact, and is likely
to be violated by operators of dimension just higher than
that permitted by our earlier arguments.  As Sikivie
noted long ago, under these circumstances, it is quite
possible that any domain walls formed at the PQ phase
transition would harmlessly disappear (\ie, before nucleosynthesis).

What are we to make of all of this?  My attitude is that
there are two, and possibly three, plausible mechanisms
for solving the strong $CP$-problem.  Thanks to
the heroic efforts of a now rather large number
of physicists, we have a real hope
of getting an experimental handle on the axion some time
soon.  The discovery of this particle would be an
extraordinary event, a connection with physics at vastly higher
energies than we have ever studied.  Perhaps, in light
of all of these criticisms, the probability of success is not
$50\%$, but even at $10\%$ the payoff seems worth it.

\chapter{NEUTRALINO COLD DARK MATTER}

\REF\kamionkowski{M. Kamionkowski, IAS Princeton preprint
IASSNS-HEP-92-37 (1992).}
\REF\wimps{I will not attempt a complete list  of references
here, but refer the reader to the collection of
articles, {\it Particle Physics and Cosmology:  Dark Matter},
M. Srednicki, ed., North-Holland (Amsterdam) 1990,
and recent overviews by Griest\refmark{\greist}
and Kamionkowski.\refmark{\kamionkowski}}
We have remarked that massive particles with weak interaction cross sections
are
ideal dark matter candidates.
It is easy to understand why this is so.\refmark{\wimps}
Let us call our WIMP $w$.
At extremely early times, when the temperature is well
above its mass, $w$ will be in thermal
equilibrium.  Equilibrium is maintained, typically, by pair
annihilation ($w + w \rightarrow q + \bar  q$, say),
and the reverse process of WIMP pair production.
At these temperatures, the $w$ density goes as
$n_w \propto T^3$.  As the
temperature drops below the $w$ mass, however, the
$w$ density falls as
$$n_w \propto (m_w T)^{3/2} \exp(-m_w T).\eqn\roughdensity$$
(The density of $f \bar f$ pairs, where $f$ denotes
light fermions, with sufficient energy to produce
$w$ pairs falls similarly).    Eventually, annihilation and
production reactions are too slow to maintain equilibrium,
and one is left with some density of $w$'s.  The density
is clearly inversely proportional to the cross section.
Detailed calculations give for the
final density, as a fraction of closure density,
$$\Omega_w h^2 \approx {3 \times 10^{-27}{\rm cm}^3{\rm sec}^{-1}
\over \sigma_A v}.\eqn\omegacrosssection$$

This is a wonderful result:  to be a suitable dark
matter candidate, the cross section should be roughly
of weak interaction strength.  In other words,
it is possible that if this particle exists, we have a chance
of discovering it in the laboratory.  Moreover, in our menu
of present theoretical ideas, we have several candidates.

What about detection of these particles?  Of course,
in some cases one can hope to observe them in accelerators. If
such particles actually make up the halo
of our galaxy, then there is a significant flux of these particles
on earth, which one might hope to observe.   One expects
that the typical velocities will be of order 220 km/sec.
Passing through quantities of matter, these particles can
scatter off nuclei.  Typical interaction rates are of order
1/kg/day for particles with principally spin-dependent,
incoherent interactions with nuclei (such as majorana particles),
while they can be orders of magnitude larger for particles
with coherent interactions.
We have mentioned that
two obvious candidates of this type are already ruled out:  Majorana and
Dirac neutrinos.  Majorana neutrinos are ruled out by the fact that if they
constitute cold dark matter, their masses are necessarily in the few GeV
range,
and this is ruled out by LEP.  Dirac neutrinos can
play the role of dark matter over a large range of mass,
since they can have associated with them
an approximately conserved quantum number (like baryon
number).  However, they are
ruled out, for masses from about $20~\gev$ to 1000~TeV by
direct searches of the type which
will be discussed below, while lower masses are ruled out by LEP.

But aside from these rather conventional(?) extensions of the
known particles of the standard model, other candidates have
emerged from studies of extensions of the standard model.  Probably
the most popular extension of the standard model is supersymmetry.
Supersymmetry near the weak scale, as you have all heard many times, is of
interest
for a variety of reasons:  it potentially solves the hierarchy problem;
it leads to a far better unification of couplings than non-supersymmetric
theories; it provides an attractive mechanism for $SU(2) \times U(1)$
breaking, and more.  Certainly one attractive feature of these
models is that they provide an excellent dark matter candidate.

Just to remind you briefly of the basics of supersymmetry, supersymmetry
is a symmetry between bosons and fermions.  If it is a symmetry of nature,
for each of the fields we currently know, there must be a bosonic or
fermionic partner as appropriate.  For example, for all of the gauge
bosons of the standard model, there must be a gauge fermion (gaugino)
in the adjoint representation of the gauge group; there must be scalar
quarks (squarks) and scalar leptons (sleptons).  From LEP and CDF we  have
lower limits on the masses of many of these particles.  If supersymmetry
has anything to do with the solution of the hierarchy problem, these particles
cannot be too heavy.  Of course, the fact that particles are not degenerate
with
their supersymmetry partners means that supersymmetry is a broken
symmetry.  Little is known about this breaking, and there are only a few
models where the breaking can be understood dynamically.  As a result,
almost all analyses to date proceed by making a set of simplifying assumptions.
First, one assumes that the particle content of the model is the minimal
one consistent with supersymmetry and the states we currently observe.
This just means that one has, in fact, a superpartner for all of the
ordinary quarks, leptons and gauge bosons, and that one has two
Higgs doublets and their ``Higgsino" partners.  One also typically makes
some assumptions about how supersymmetry is broken, \ie, about
the ``soft breaking" parameters, the masses and (superrenormalizable)
couplings of the superfields. In particular, most workers assume that
at the scale of unification, the masses of all of the scalar fields are
identical, and that there is a proportionality between soft breaking
cubic couplings and Yukawa couplings.  With these assumptions, one
has a model specified by a relatively small number of parameters:
the top quark Yukawa coupling, $h_t$, the gaugino masses, $m_{1/2}$,
the common squark and slepton masses, $m_o^2$, a supersymmetric
Higgs mass term, $\mu$, and quantities $A$ and $B$ which describe
the relations between the Yukawa couplings and the cubic terms
(and a certain term involving the Higgs masses).  The low energy
parameters are then determined by running the masses and couplings
down to low energies using the renormalization group (usually after
having determined the unification scale and coupling by the standard
renormalization group analysis.)  One of the exciting
features of these models is that, for a range of parameters, this running
leads to a negative mass for the Higgs field which couples to the top
quark, which triggers $SU(2) \times U(1)$ breaking.

\REF\nonr{S. Dimopoulos, R. Smailzedeh, L. Hall,
J.P. Merlo and G. Starkman, Phys. Rev. {\bf D41}
(1990) 2099.}
There is one other, critical assumption which is usually made in these models.
This is that the models should possess a discrete symmetry known as $R$-parity,
under which all ordinary fields are neutral (including the two Higgs doublets)
while all of the new superfields change sign.  This is necessary to forbid
vertices  which would lead to rapid proton decay.
(Alternatives to $R$-parity will not be considered
here.\refmark{\nonr})   This symmetry has an immediate consequence:
the lightest of the new particles is stable; it is this particle which is a
candidate for dark matter.  Which of these states it is depends on the
parameters of the model.  One possibility which has been ruled
out is the ``sneutrino," the scalar partner of the neutrino.
 From LEP we know that this state, if it exists, has a mass larger than about
$42~\gev$.  $\tilde \nu$ annihilation would proceed through
``zino" exchange. But this interaction is rather strong, and as a
result, the
$\tilde \nu$ abundance is too small to make it a dark matter candidate.

\REF\neutralinoreviews{L. Roszkowski, University of Michigan preprint
UM-TH-92-06 (1992).}
\FIG\neutralinos{A sampling of neutralino results, illustrating
that for a significant part of the parameter space, neutralinos
can give $\Omega=1$.}
For a good part of the parameter space, however, the LSP is the ``neutralino,"
a linear combination of the Higgsino, the ``photino" and the ``zino."
We will denote this state by $\chi$.  In the MSSM, for a
significant range of parameters,
one finds that $\chi$ is a good dark matter candidate.
This problem has been studied by many authors, and I will not attempt a
complete survey here.\refmark{\greist,\kamionkowski,
\neutralinoreviews}
For example, in fig.~\neutralinos\
one sees, for somewhat different assumptions about the parameters,
that $\Omega \sim 1$ for a significant range of parameter space.
Based on this, many have argued that if supersymmetry
exists, the LSP is the dark matter.

\REF\griestr{K. Griest and L. Roszkowski, Phys. Rev. {\bf D46} (1992)
3309.}
\midinsert
   \tenpoint \baselineskip=12pt   \narrower
\vskip7.5cm
\vskip6pt\noindent
{\bf Fig.~\neutralinos.}\enskip
{A sampling of neutralino results, illustrating
that for a significant part of the parameter space, neutralinos
can give $\Omega=1$ (from ref.~\griestr).}
\endinsert

\REF\sophisticated{See, for example, M. Drees and M.M. Nojiri,
Phys. Rev. {\bf D47} (1993) 376;
S. Mizuta and M. Yamaguchi, Phys. Lett. {\bf B298} (1993) 120;
J. McDonald, K.A. Olive and M. Srednicki, Phys. Lett. {\bf B283}
(1992) 80; M. Drees, G. Jungman, M. Kamion\-kowski
and M.M. Nojiri, Princeton IAS preprint IASSNS-HEP-93/37 (1993).}
\REF\roberts{R.G. Roberts and L. Roszkowski, Phys. Lett. {\bf B309} (1993)
329.}
Over the past year, there have been two significant developments.
First, the problem of neutralino annihilation and detection has been
studied with great sophistication and power.\refmark{\sophisticated}  Much of
the
MSSM parameter space has been ruled out; gaugino-like
LSP's have been seen to be
favored.  Second, it has been appreciated that
with some assumptions, one can even use the hypothesis of neutralino
dark matter to constrain the supersymmetry parameters.  Typically,
one doesn't want masses of squarks, etc., to be too small, since in that case
the annihilation cross sections are too large.  Roberts and
Roszkowski,\refmark{\roberts}
for example, find the following sorts of parameters (assuming, in addition
to unification of gauge couplings, $SU(5)$-type unification of $m_b$
and $m_{\tau}$.):
$$60<m_{\chi} <200\ \gev ~~~~~
150<m_{\chi^{\pm}} <300\ \gev ~~~~~
200<m_{\tilde l} <500\ \gev \eqn\rrlimits$$
$$250<m_{\tilde q} <850\ \gev ~~~~~
350<m_{\tilde g} <900\ \gev.$$
The heavy Higgs fields have masses of order 250--700~GeV.
Considerations of naturalness probably prefer the lower values of these
masses.

\FIG\directdetection{Figure from ref.~\roberts\ indicating the
fraction of the
MSSM parameter space which can be ruled out by direct
detection experiments.}
\REF\cpagroup{T. Shutt \etal, Phys. Rev. Lett. {\bf 69} (1992) 3425.}
What are the prospects for detecting this dark matter?
If nature is supersymmetric at low energies,
we are certain (given sufficient cooperation from the taxpayers) to
find it in accelerators in the not too distant future.  But can we hope to
see the dark matter directly?  This turns out not to be as easy as in the
case of neutrinos, mentioned above.  The problem is that $\chi$ is a Majorana
particle, with only axial couplings to ordinary quarks at low momenta.
As a result, one does not have the sort of coherent effects which permit
heavy neutrino detection.  Actually, the situation is more subtle than this.
Once one considers couplings to gluons through loops, coherent couplings
do arise.  Even then, the event rates are small;
typically one is talking of rates
of order 1 /kg/day  This is problematic, given that backgrounds
from photons are typically of order 0.5/kg/day/keV.  Still, there are
many groups trying to search for these events.  Recently,
the feasibility of measuring the phonon signal
from nuclear recoil has been demonstrated.\refmark{\cpagroup}
A larger pilot experiment
is currently in progress (involving LBL,CPA, UCSB, Stanford
and San Francisco State; the studies are being performed
at SLAC, using $1/2$ kg of $^{76}$Ge, brought
over from the FSU in a suitcase).  This gives significantly greater
sensitivity,
since
photons deposit 33\% of their energy in ionization, and 66\%
in phonons, while recoiling nuclei deposit 90\% of their energy in phonons.
As a result, one can examine a significant part of the neutralino parameter
space, as indicated in fig.~\directdetection.

\topinsert
   \tenpoint \baselineskip=12pt   \narrower
\vskip10cm
\vskip6pt\noindent
{\bf Fig.~\directdetection.}\enskip
{Figure from ref.~\greist\ indicating the fraction of the
MSSM parameter space which can be ruled out by direct
detection experiments.}
\endinsert

\REF\gould{A. Gould, Ap.~J. {\bf 3211} (1987) 571.}
An alternative method of searching for neutralinos from the halo
is also being actively pursued.  Neutralinos (and similar WIMP's)
can be captured in the earth and the sun, where they can subsequently
annihilate.  Since the sun and the earth have been around for a long
time, the total numbers of captured particles can be substantial.
Capture occurs when a neutralino scatters elastically on a particle in the
sun, emerging with a velocity smaller than the escape velocity.
(The most thorough calculations of the capture rate
are probably those by Gould.\refmark{\gould})
The typical velocities of these particles
are of order 200~km/sec.  Because of the earth's low escape velocity,
and because most of the nuclei in the earth are spinless,
capture is rare for particles with mass greater than about $80~\gev$
(for pure gauginos).
In the sun, capture is much more probable.  The subsequent annihilation
of these WIMP's in the sun leads to production
of $\nu$'s.  For neutralinos, these
$\nu$'s are produced principally in cascade decays
of heavy quarks; most are $\nu_{\mu}$'s,
with typical energies  20--30 GeV
To date, the best experimental results come from IMB and
Kamiokande.   There
are significant backgrounds from cosmic rays.  Kamiokande, I understand,
is currently doing a careful analysis.  MACRO, in the Gran Sasso,
is currently running; it covers an area something of the order of
a football field in size.  Future experiments include
DUMAND and AMANDA.  This last experiment
involves looking for $\nu$'s by sinking an array of phototubes
in the south pole ice; some pilot experiments
have already been done.  In principle, this experiment
should cover an area about $100$ times as big as that of MACRO.
Eventually one can hope to rule out a significant part of the
MSSM parameter space.

\FIG\indirect{Some of the parameter space which can
be covered by indirect searches, from ref. \kamionkowski.}
What sort of uncertainties exist in these
experiments?  They are surprisingly small.  The local density
is known to within
a factor of two.  Uncertainties in the elastic cross sections
are of a similar order; perhaps slightly larger.
Some sense of what part of the parameter space can be ruled out (or discovered)
by these types of experiments is indicated in fig.~\indirect.

\midinsert
   \tenpoint \baselineskip=12pt   \narrower
\vskip9cm
\vskip6pt\centerline{%
{\bf Fig.~\indirect.}\enskip
Some of the parameter space which can be covered by indirect
searches.}
\endinsert

Finally, I would like to turn to two more theoretical questions.
Most of the analyses which I have described involve the MSSM.
They make quite specific assumptions about the mass spectrum.
Just how plausible are these assumptions?   And how important are
they to these analyses?

\REF\ibanezlust{L. Ibanez and D. Lust, Nucl. Phys. {\bf B382}
(1992) 305.}
\REF\vadim{V.S. Kaplunovsky and J. Louis,
 Phys.~Lett. {\bf B306} (1993) 269.}
\REF\nirseiberg{Y. Nir and N. Seiberg, Phys. Lett. {\bf B309} (1993) 337.}
\REF\dinenelson{M. Dine and A.E. Nelson,
Phys. Rev. {\bf D48} (1993) 1277.}
In the absence of data, the answer to the first question depends a good
deal on personal prejudice.  For example, consider the assumptions that squarks
and sleptons are degenerate at the highest energy scale,
and that the gaugino masses are identical at this scale.  Some assumption
like this, apart from its plausibility, is essential in order to understand the
absence of flavor changing neutral currents.
In virtually all models which have been studied to date, however, all
of these masses are parameters, which are not constrained by any
principles of symmetry.  Recently, a variety of scenarios have been
proposed which might give rise to the phenomenologically required degree
of degeneracy and proportionality.  In string theory (which, correct or not,
is the only theory of gravity we currently possess), in a generic sense
there is no degeneracy.\refmark{\ibanezlust}  However, it has recently been
recognized that there are circumstances under which a significant degree of
degeneracy does arise (though perhaps not quite large enough to explain the
absence of FCNC's).\refmark{\vadim}  An alternative approach is to assume
that there are some underlying flavor symmetries, broken at a high energy
scale.\refmark{\dkl}
Such an approach predicts a certain degree of degeneracy, but
no degeneracy between squarks and sleptons, or
between squarks in the first two and the third generations,
or  possibly no degeneracy at all.\refmark{\nirseiberg}
Finally, it should perhaps
be mentioned that models exist in which supersymmetry is dynamically
broken at low energies.\refmark{\dinenelson}
In these models, the lightest superparticle is
a very light gravitino; there is no conventional cold dark matter candidate
in the neutralino sector.   These models predict even less degeneracy,
but their cosmological implications are different, and I will mention
them only briefly towards the end, when I discuss various more
exotic possibilities.
Some work attempting to relax the usual MSSM assumptions
already exists,\refmark{\griestr} but further efforts, perhaps
motivated by some of the ideas outlined above, would
be worthwhile.

\chapter{BARYOGENESIS}

\REF\cknreview{A.G. Cohen, D.B. Kaplan and A.E. Nelson,
UC San Diego preprint UCSD-PTH-93-02 (1993).}
In the last few years, it has become clear that the
standard model violates baryon number significantly
at high temperatures.  This opens the possibility that the
electroweak phase transition, if it is first order, could
be the origin of the baryon asymmetry.  Several mechanisms
have been suggested which could lead to the
asymmetry. The
subject has recently been the subject of an excellent
review,\refmark{\cknreview} so I will only mention here
two recent developments, and areas where further work is
needed.

\REF\arnold{P. Arnold and O. Espinosa, Phys. Rev. {\bf D47} (1993) 3546.}
\REF\bagnasco{M. Dine and J. Bagnasco, Phys. Lett. {\bf B303} (1993) 308.}
\REF\cknthree{A. Cohen, D. Kaplan, and A. Nelson,
Nucl. Phys. {\bf B263} 1991) 86.}
In order to obtain a departure from equilibrium, it is
necessary that the phase transition be first order.
Over the last year, there has been a great deal
of work attempting to understand the phase transition.
Perhaps the most thorough of these studies is that due
to Arnold and Espinosa.\refmark{\arnold}
 (A somewhat simpler version
of this analysis, which yields the largest contribution,
is due to myself and my student J. Bagnasco\refmark{\bagnasco}).
This work
suggests that, for the relevant range of parameters,
perturbation theory is not too bad a guide.  Still, many
serious questions have been raised about the validity
of the perturbative analyses, and further work is necessary.
There has been a great deal of other work on the phase
transition as well, and controversy still seems to rein
about such questions as:  how strongly first
order is the transition?  how do bubbles propagate?

\REF\farrar{G. Farrar and M. Shaposhnikov, Phys. Rev. Lett.
{\bf 70} (1993) 2833; CERN-TH-6732-93.}
\REF\dinedpf{M. Dine, in {\it The Vancouver Meeting,
Proceedings of the DPF Meetings,
Vancouver}, August 18--22, 1991, D. Axen and D. Bryman, eds.
(World Scientific, Singapore, 1992).}
Most discussions of the subject of electroweak baryogenesis
begin with the remark that there is not enough $CP$-violation
in the minimal standard model (MSM) to give the observed
asymmetry.
Since three generations are required to obtain $CP$-violation,
one expects suppression by mixing angles and quark masses;
even before one begins this yields a number like $10^{-20}$.
Recently, however, Farrar and Shaposhnikov have argued
that the suppression may not be nearly so severe.  The argument
involves careful treatment of quasi particle
excitations in the plasma and resonant phenomena similar to
the MSW effect.  I think it is safe to say that the verdict is not yet
in.  Even if the suppression is not at great as one might have
imagined, as these authors note, there are still a number
of obstacles to obtaining a reasonable asymmetry in the
MSM.  First, a perturbative
analysis gives that the baryon asymmetry is washed out unless
the Higgs mass is less than about $35$ \gev.  The recent analyses
referred to above indicate that higher order corrections give
only a small change in this limit.  Against this, one might
argue that for heavy enough Higgs, the finite temperature
perturbation theory is not under control at all.  Second, there
is the question of the rate.  In ref.~%
\cknthree, similar dynamics in an extended model gave a maximum
asymmetry of about $10^{-4}$.  In comparing the
analysis of ref.~\farrar\ with this,
one sees that in the MSM one must pay at least a factor
of $10^{-5}$ for mixing angles, and a factor of $m_s /T \sim
10^{-3}$.   While their analysis differs somewhat, the
main source of  the larger answer seems to
be in the choice of the baryon number violating rate.
Such a large
value is not implausible (indeed I have argued for it
elsewhere\refmark{\dinedpf}).

\REF\dortmund{J. Baacke and S. Junker, Dortmund preprint
DO-TH-93/15.}
\REF\carson{L. Carson, X. Li, L. McLerran and R.-T. Wang, Phys.
Rev. {\bf D42} (1990) 2127.}
\REF\comelli{D. Comelli, M. Pietroni and A. Riotto,
DFPD-93-TH-26 (1993).}
\REF\schmidt{C.R. Schmidt and M. Peskin,
Phys. Rev. Lett. {\bf 69} (1992) 410; C.R. Schmidt,  Phys. Lett.
{\bf B293} (1992) 111; B. Grzadkowski and
J.F. Gunion, Phys. Lett. {\bf B287}
(1992) 237;
C.J.C. Im, G.L. Kane and P.J. Malde, University of Michigan preprint
UM-TH-92-27 (1992).}
\REF\gunion{B. Grzadkowski and J.F. Gunion, Phys. Lett. {\bf B294}
(1992) 361.}
There are some other developments which I would like to
briefly mention:
\item{1.}   The baryon number violating rate in the broken
phase has been recalculated,\refmark{\dortmund} yielding
a larger value than earlier calculations.\refmark{\carson}
This is suggestive
that the rate in the unbroken phase may be larger than
previously assumed.
\item{2.}  There has been an interesting proposal for how
a suitable asymmetry might arise in a theory like the MSSM
where $CP$-violating phases are small.\refmark{\comelli}
The point is that
in many theories there is a range of parameters for which
$CP$ is spontaneously violated at high temperatures.
One might expect that this would lead to equal
numbers of regions with one
sign or the other of the baryon number, and that there would
be no net asymmetry.  However, because the
bubble nucleation rate is exponentially sensitive to
a large, three-dimensional tunneling action,
small fractional changes in this action due to a small, $CP$-violating
asymmetry can significantly bias the rate of bubble formation.
\item{3.}  Motivated by the possibility of electroweak baryogenesis,
a number of authors have begun to study the observability
of various $CP$-violating phenomena in different experimental
environments, including the SSC\refmark{\schmidt}
and NLC.\refmark{\gunion}

Despite the progress of the last few years, as the confusion
about baryogenesis in the MSM illustrates,
it remains important to have improved calculations of
baryon number violating rates, particularly in the unbroken
phase, and to better understand the electroweak
phase transition and specific mechanisms for
producing the asymmetry.

\chapter{EXOTICA}

\REF\preskill{J. Preskill, S.P. Trivedi, F. Wilczek and M.
Wise, Nucl. Phys. {\bf B363} (1991) 207.}
\REF\senjanovic{B. Ral and G. Senjanovic,
ICTP preprint IC-92-414.}
\REF\banksetal{T. Banks, D.B. Kaplan and A.E. Nelson,
UC San Diego preprint UCSD/PTH 93-26 (1992).}
\REF\masiero{S.A. Bonometto, F. Gabbiani and A. Masiero,
Padua preprint DFDD-93-TH-33 (1993).}
Finally, I  would like to turn to a number of more exotic
problems, somewhat further removed from direct observation.
Necessarily, because time is limited, these choices reflect more
personal interests.
Cosmology, in principal, can provide many interesting constraints
on model building.  If we assume that at temperatures of order,
say, the weak scale and above, the universe was always
in thermal equilibrium, these constraints can be quite
severe.  We have already mentioned one example:  gravitinos
lead to trouble, unless the reheat temperature after inflation
is sufficiently low.  Let me mention some examples which
have been of relevance recently:
\pointbegin
 Domain walls.  In models with spontaneously broken
discrete symmetries, domain walls form.  If they do not somehow
decay, these come to dominate the energy density of the
universe, and are unacceptable.  A simple example of this
problem is provided by a model with two Higgs doublets,
$\phi_1$ and $\phi_2$ (non-supersymmetric).  In such a model,
one usually imposes a discrete symmetry which prevents flavor
changing neutral currents.  Because of this symmetry,
the theory would appear to have a pair of degenerate
vacua at tree level.  However, Preskill et
al\refmark{\preskill} pointed out that once QCD effects are taken
into account, these vacua are not quite degenerate; they are
split by an amount of order $m_{\pi}^2 f_{\pi}^2$.
At first sight, this would appear irrelevant; it is to
be compared with $m_W^4$.    However, because the
expansion of the universe goes as $H\sim {m_W^2 / M_P}$,
there is plenty of time for the domain walls to collapse.
This suggests a more general point which was first raised, to my
knowledge, by Sikivie\refmark{\sikiviewalls}
in the context of axion physics.  Suppose
one has an approximate (accidental) discrete symmetry,
broken by dimension 5 operators.  Then these operators
can also lead to collapse of domain walls.  Thus it is not
clear that domain walls are such a problem for model building.
This idea was recently revived by Ral and Senjanovic\refmark{
\senjanovic} in a different
context, and by myself and A. Nelson in the framework of
models of dynamical supersymmetry
breaking, where accidental discrete
symmetries (\ie, symmetries which are accidental
consequences of gauge invariance and renormalizability)
are common.\refmark{\dinenelson}
\point
Problems with an earlier model:  In my discussion of axions,
I described a class of models in which discrete symmetries
gave rise to an accidental Peccei-Quinn symmetry good
enough to solve the strong $CP$-problem.  However, I did
not go into great detail about these models, and I neglected
to point out another cosmological problem that these raise.
One of the attractive features of these models turns out
to be that the scale $f_a$ is automatically of the desired
order of magnitude,
$f_a \sim m_W m_P$.
However, there is another side to this good feature.  In these
theories, the PQ symmetry is broken by the vev of a field
whose potential has a characteristic curvature of order
$m_{3/2}$.  How does the PQ-violating phase transition
arise in this model?  If the field, $S$, is in thermal equilibrium
at high temperature, than near the origin its potential has
a curvature of order $T^2$.  But this means that the field
gets hung up for a while in a false vacuum.  Eventually,
it rolls to its minimum, but at the minimum all of the fields
to which it couples are very massive, so it can only dissipate
its energy with great difficulty.  This is potentially
a catastrophe.  While this problem is worthy of further investigation,
one solution is to suppose that the field does not ``start"
near its minimum.  In this case, when it starts to oscillate
it carries only a tiny fraction of the total energy density,
and this is still true when it decays.
\point
Axinos:
Related to the problem described above is a perhaps more
general set of issues which have been extensively discussed
in the literature.  In supersymmetric theories in which the strong
$CP$-problem is solved by an axion, the axion will have a scalar
and a spinor superpartner.  The spinor is referred to as the
axino, the scalar as the saxion.  Both of these particles are,
like the axion, extremely weakly interacting, with interaction
strengths proportional to $1/f_a$.  Thus they have
potentially significant phenomenological implications.  It is generally
agreed that the saxion will have a weak-interaction type mass
after supersymmetry breaking.  About the axino, the literature
is more varied, with many claims that this particle can be quite
light, with a mass of order $m_{3/2}^2 /f_a \sim\ $keV.
This would have implications, for example, for the stability
of neutralinos, and thus for their role as dark matter; they
might well decay on cosmologically interesting time scales
(if their decays are due to operators of dimension $5$, for example,
their lifetimes could well be of order seconds; if higher dimensions,
they could be of order the age of the universe).
These axinos, in turn, could play the role of ``warm" dark
matter.  In fact, Masiero \etal\ have recently pointed out
that they could provide a form of HCDM, with one component
arising from neutralino decay, another from the simple equilibration
of these particles.\refmark{\masiero}
\REF\wilczek{K. Rajagopal, M.S. Turner and F. Wilczek,
Nucl. Phys. {\bf B358} (1991) 447.}
\REF\axinomass{See, \eg, T. Goto and
M. Yamaguchi, Phys. Lett.
{\bf B276} (1992) 103.}

\item{}\hskip12pt
It should be noted, however,
that these axinos are likely to have weak interaction size
masses.  In the context of particular models, this has been
discussed before.\refmark{\axinomass}
In general, a supersymmetric mass term for these fields
can appear in the effective lagrangian with a coefficient
of order $m_W$ ($m_{3/2}$) by essentially the same mechanisms
which have been discussed for the Higgs particles.
(For a nice review of this latter problem, see ref.~\vadim.)
If $\phi = \tilde s + ia + \theta \tilde a + \dots$ denotes the
axion superfield, and $Z$ is some hidden sector field responsible
for supersymmetry breaking, the operator
$$\int d^4 \theta Z^{\dagger}(\phi+\phi^{\dagger})^2
\eqn\axinomassop$$
would give a mass to the axino of order $m_{3/2}$;
in the context of supergravity, other terms can appear as well.
It is difficult to suppress these operators by any symmetry, since
such a symmetry would also forbid gaugino masses.
If they are heavy, the cosmology of the axino and
saxino fields is potentially catastrophic, since these fields drop
out of equilibrium while they are still relativistic, and easily
provide far too much matter today.\refmark{\wilczek}
This problem, however,
may be solved if inflation occurs below the Peccei-Quinn
scale, at least if the reheating temperature is low enough
that not too many of these particles are produced later
(this constraint is analogous to constraints on gravitinos).
\point
Cosmological constraints on models with dynamical supersymmetry
breaking.  Recently, Banks \etal\refmark{\banksetal}
have surveyed a number of
issues in models with dynamical supersymmetry breaking.
They point out that, making again the sort of cosmological
assumptions I have described above, many supersymmetry
breaking scenarios have trouble.  In particular, gluino
condensation scenarios for string theory run into difficulties
analogous to some of those I have described above.  One has,
at early times, weakly coupled fields which have no reason to
sit near the minima of their potentials.  As a result, huge
amounts of energy are stored in them, and it is difficult to
dissipate this energy.  Instead, one is forced to a particular
class of models in which supersymmetry is broken strongly.
These models have their own special difficulties, in particular
in generating gaugino masses.
\point
Still other features of models with dynamical supersymmetry breaking.
Recently, Ann Nelson and I have considered a class of models
in which supersymmetry is broken dynamically at TeV
energies.\refmark{\dinenelson}
The only feature of these models I want to mention is that in
these theories, the neutralinos and similar particles can decay
with weak interaction lifetimes to gravitinos; the gravitinos, in
turn, are very light (typically with eV-ish masses).  These
particles in themselves are a potential problem.  They are like
a fourth generation of neutrinos (with two helicity states)
and can spoil nucleosynthesis.  These models have other problems
as well, which have been alluded to earlier.  They can give rise
to domain walls (though the particular example we studied does
not) which must be removed by the higher dimension operator
mechanism I described earlier.  Also, these models often contain
massive, stable particles associated with supersymmetry breaking.
These are probably not suitable dark matter candidates; in fact,
there are likely to be far too many of them unless higher dimension
operators render them unstable.

Earlier, I stressed that all of these problems presume that
we can simply run the clock backwards on the big bang.
This is not necessarily so.  Since we now know that the
baryon asymmetry may have been created rather late,
one can contemplate the possibility that there was a severe
departure from thermal equilibrium at temperatures far
below any GUT scale.  The most extreme possibility, which
has been developed recently by
Knox and Turner is that inflation itself
occurred at the weak scale.\refmark{\knox}
Before the advent of electroweak
baryogenesis, this would have been deemed impossible;
there would be no way to generate the observed asymmetry.
Now, it appears that if the final temperature after inflation
is of order the weak scale, it is possible to generate the asymmetry.
It is probably fair to say that no very attractive model of this
type exists.  However, many would argue that the same is
true of inflation at any scale, so this possibility should
be born in mind.

Let me close by mentioning another class of puzzles.
I have, so far, been almost taking for granted that
$\Omega=1$.  But there is a problem with such a value
of $\Omega$:  the age of the universe.  We know from
a variety of sources (long-lived radioisotopes,
oldest stars, cooling of white
dwarfs) that the age of the universe
is almost certainly greater than about $12$ billion years.
On the other hand, most measures of $h$ give $H \approx 80\ {\rm km~%
sec}^{-1}~{\rm Mpc}^{-1}$.  Assuming $\Omega=1$, this gives
for the expansion age of the universe $t=8$ billion  years.
Some measurements of $H$ give values around
$50$, giving something closer to the $13$ billion
years or so inferred from other measurements.
Many astrophysicists believe that
the lower value will eventually be seen to
be the correct one, but one should keep in mind
that this is not necessarily the case. In principle, this question will
be settled eventually by HST.

\REF\turnerreview{For a discussion and refs., see, for example,
M. Turner, FERMILAB-CONF-92/382-A.}
One model which has been proposed to explain a larger value of $H$
has $\Omega$=0.2--0.3 in cold dark matter or baryons,
while the universe is flat as a result of a cosmological constant.
Such a model can provide for growth of structure while
at the same time accommodating the larger values for the
age of the universe.\refmark{\turnerreview}

\REF\carlson{D. Garretson and E. Carlson, Harvard University preprint
HUTP-93-A020 (1993).}
\REF\weinberglambda{S. Weinberg,  Phys. Rev. Lett.
{\bf 59} (1987) 2607.}
In any
case, this dilemma has created interest in models with
a non-vanishing cosmological constant.
Of course, it is one of the great mysteries of particle physics
why the cosmological constant is small.  It is perhaps doubly
mysterious why $\Lambda$ might take on just the value
of cosmological interest.  Recently, Carlson and Garretson have
noted that this value might not be unreasonable in a theory
which could solve the cosmological constant problem, but
in their scenario such a value is not in any sense
preferred (the cosmological constant is just
a function of some mass scales which are adjustable).\refmark{\carlson}
My own belief, if the cosmological constant is
non-zero, is that some sort of anthropic explanation is called
for (perhaps the cosmological constant takes different values
in different parts of the universe, and for some  reason -- which
I certainly don't claim to know -- we can only exist in a
universe with such a small cosmological constant).\refmark{\weinberglambda}

\bigskip
\centerline{ACKNOWLEDGMENTS}
\smallskip
I would like to thank K. Griest, M. Kamionkowski, L. Krauss,
J. Primack, M. Peskin, M. Srednicki, K. van Bibber,
and H.E. Haber for educating me
about many of the issues discussed here.  They are not,
of course, responsible for any errors or omissions.
This work is supported in part by the U.S. Department of Energy.

\refout
\end